\documentclass[twocolumn,twoside,preprintnumbers,amsmath,amssymb,pacs]{revtex4}
\usepackage{epsfig}
\usepackage{graphicx}
\usepackage{dcolumn}
\usepackage{bm}

\usepackage{pslatex}

 \newcommand\beq{\begin{equation}}
 \newcommand\eeq{\end{equation}}
 \newcommand\beqn{\begin{eqnarray}}
 \newcommand\eeqn{\end{eqnarray}}
 \newcommand{\la}{\langle}  
 \newcommand{\ra}{\rangle}
 
\def\fm{\,\mbox{fm}}
\def\GeV{\,\mbox{GeV}}

 \def\Pom{{ I\!\!P}}

\def\im{{\rm Im}}

\def\fm{\,\mbox{fm}}
\def\GeV{\,\mbox{GeV}}

\def\Pom{{\bf I\!P}}

\begin{document}
\hspace*{13cm}
\begin{minipage}{4cm}
\end{minipage}
\bigskip

\title{{\Large Handedness of direct photons }}

\author{B. Z. Kopeliovich, A. H. Rezaeian, Ivan Schmidt}

\affiliation{Departamento de F\'\i sica y Centro de Estudios Subat\'omicos,
Universidad T\'ecnica Federico Santa Mar\'\i a, Casilla 110-V, Valpara\'iso,
Chile}



\begin{abstract}



The azimuthal asymmetry of direct photons originating from primary
hard scatterings between partons is calculated. This can be accounted
for by the inclusion of the color dipole orientation, which is
sensitive to the rapid variation of the nuclear profile. To this end 
we introduce the dipole orientation within the saturation model of
Golec-Biernat and W\"usthoff, while preserving all its features 
at the cross-section level.  We show that the direct photon
elliptic anisotropy v2 coming from this mechanism changes sign and
becomes negative for peripheral collisions, albeit it is quite
small for nuclear collisions at the RHIC energy.

\end{abstract}

\maketitle

\setcounter{page}{1}

\section{Introduction}

Direct photons can be a powerful probe of the underlying dynamics of
the initial state of matter created in heavy ion collisions, since
they interact with the medium only electromagnetically and therefore
provide a baseline for the interpretation of jet-quenching
models. There are several sources for direct photons, including prompt
photons produced from initial hard scattering, thermal radiation from
the hot medium and photons induced by final state interactions with
the medium.

Unfortunately, the advantages of direct photons as a clean signature of the
initial state of matter created in heavy ion collisions are offset by
large backgrounds coming from hadronic decays, which should be
extracted. The PHENIX collaboration at RHIC has recently reported some
results of the measurement of direct photon production \cite{phe,phe2}
, which has been also subject of studies in several theoretical papers
\cite{us-1,us-2,nd,nd-n,pd,la}. 

A novel mechanism which produces an azimuthal asymmetry coming from the
reaction's initial conditions was introduced in Refs.~\cite{us-1,us-2}.
This is in contrast with the standard approaches where the azimuthal
asymmetry is only associated with the properties of the medium created in
the final state. In our approach, the main source of the azimuthal
asymmetry originates from the sensitivity of parton multiple interactions
to the steep variation of the nuclear density at the edge of the nuclei,
which correlates with the color dipole orientation. In order to introduce
a dependence on dipole orientation, we extend the model of Golec-Biernat
and W\"usthoff \cite{kst} for the total dipole cross section to the partial
dipole-nucleon amplitude. To do that we assume that the two gluons in the
Pomeron are not correlated.

\section{Photon radiation in the colour dipole formalism} 

\begin{figure}[!t]
       \centerline{\includegraphics[width=6 cm] {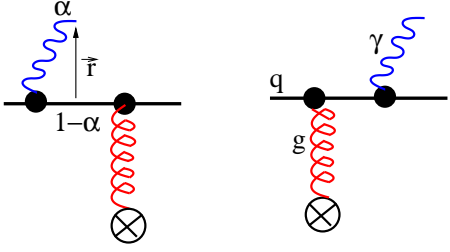}} \caption{
       Direct photons production in the target rest frame. The photon
       can be also radiated before the quark hits the target. Both
       diagrams are important. The dipole cross-section appears since
       the quark is displaced in impact parameter plane after
       radiation of photon. The anti-quark enters after taking the
       complex conjugate of the amplitude, which is not shown here.
       \label{f1} }
\end{figure}

Radiation of direct photons in the target rest frame should be treated as
electromagnetic bremsstrahlung by a quark interacting with the target. In
the light-cone dipole approach the transverse momentum distribution of
photon bremsstrahlung by a quark propagating and interacting with a target $t$
(nucleon, $t=N$, or nucleus, $t=A$) at impact parameter $\vec b$, as
calculated from the diagrams in Fig.~\ref{f1}, can be written in the
factorized form \cite{us-1,kst1},
 \begin{eqnarray}
&&\frac{d \sigma^{\gamma}(qt\to\gamma X)}
{d(ln \alpha)\,d^{2}{p}\,d^{2}{b}}(b,p,\alpha)=\frac{1}{(2\pi)^{2}}
\sum_{in,f}\int d^{2}{r}_{1}d^{2}{r}_{2}
e^{i \vec{p}\cdot
(\vec{r}_{1}-\vec{r}_{2})}\nonumber\\
&\times&\phi^{\star}_{\gamma q}(\alpha, \vec{r}_{1})
\phi_{\gamma q}(\alpha, \vec{r}_{2})
F_t(\vec b,\alpha\vec{r}_{1},\alpha\vec{r}_{2},x), \label{m1}
\end{eqnarray} where $\vec p$ and $\alpha=p_\gamma^+/p_q^+$ are the
transverse and fractional light-cone (LC) momenta of the radiated
photon and $\phi_{\gamma q}(\alpha,\vec{r})$ is the LC
distribution amplitude for the $q\gamma$ Fock component with
transverse separation $\vec r$, 
\beq
\phi_{\gamma q}(\alpha,\vec r_T)=
\frac{\sqrt{\alpha_{em}}}{2\,\pi}\,
\chi_f\,\widehat O\,\chi_i\,K_0(m_q \alpha r_T)
\label{dylcl}
\eeq
Here $\chi_{i,f}$ are the spinors of the initial and final quarks and
$K_0(x)$ is the modified Bessel function.  The operators $\widehat O$
have the form,
\beq
\widehat O = i\,m_f\alpha^2\,
\vec {e^*}\cdot (\vec n\times\vec\sigma)\,
 + \alpha\,\vec {e^*}\cdot (\vec\sigma\times\vec\nabla)
-i(2-\alpha)\,\vec {e^*}\cdot \vec\nabla\ ,
\eeq
where $\vec e$ is the polarization vector of the photon, $\vec n$ is a
unit vector along the projectile momentum, and $\vec\nabla$ acts on
$\vec r_T$. The parameter $m_q$ is the effective quark mass, which is
in fact an infra-red cutoff parameter  $m_q\approx 0.2\GeV$.

In equation (\ref{m1}) the effective partial amplitude $F_t(\vec
b,\alpha\vec{r}_{1},\alpha\vec{r}_{2},x)$ is a linear combination of $\bar qq$
dipole partial amplitudes at impact parameter $b$,
 \begin{eqnarray}
F_t(\vec b,\alpha\vec{r}_{1},\alpha\vec{r}_{2},x)&=&
\im\Bigl[
f^t_{q\bar{q}}(\vec b,\alpha \vec r_{1},x)+
f^t_{q\bar{q}}(\vec b,\alpha \vec r_{2},x)
\nonumber\\
&-&
f^t_{q\bar{q}}(\vec b,\alpha(\vec{r}_{1}-
\vec{r}_{2}), x)\Bigr]\,,
\label{sig}
\end{eqnarray}
 where $x$ is Bjorken variable of the target gluons.
The partial elastic amplitude $f^{A}_{q\bar{q}}$ can be
written, in the eikonal form, in terms of the dipole
elastic amplitude $f^{N}_{q\bar{q}}$ of a $\bar{q}q$ dipole
colliding with a proton at impact parameter $\vec{b}$,
\begin{eqnarray}
\text{Im}f^{A}_{q\bar{q}}(b,\vec{r},x)&=&1-\Big[1-\frac{1}{A}\int d^{2}\vec{s}~ \text{Im}f^{N}_{q\bar{q}}
(\vec{s},\vec{r},x)T_{A}(\vec{b}+
\vec{s})\Big]^{A}\nonumber\\
&\approx&1-\exp[-\int d^{2}\vec{s}~ \text{Im}f^{N}_{q\bar{q}}(\vec{s},\vec{r},x)T_{A}(\vec{b}+\vec{s})].
\label{eik}\nonumber\\
\end{eqnarray}

The hadronic cross section can be obtained by a convolution of the partonic cross section Eq.~(\ref{m1}) with a proton structure function 
$F_{2}^{p}(x,Q)$ \cite{amir2},
\begin{equation}
\frac{d\sigma^{\gamma}(pt\to\gamma X)}{dx_{F}d^{2}\vec{p}_{T}d^{2}\vec{b}}=F_{2}^{p}\otimes   
\frac{d \sigma^{\gamma}(qt\to \gamma X)}{d(ln \alpha)d^{2}\vec{p}_{T}d^{2}\vec{b}}, \label{con1}
\end{equation}
where $x_F$ denotes the Feynman variable. We take the parametrization
for the proton structure function given in Ref.~\cite{ps}.

We have recently shown that in this framework one can obtain a good
description of the cross section for prompt photon production data for
proton-proton (pp) collisions at RHIC and Tevatron energies
\cite{amir2}. Notice also that in contrast to the parton model, in this approach  
neither K-factor (NLO corrections), nor higher twist corrections are
to be added. No quark-to-photon fragmentation function is needed
either. Indeed, the phenomenological dipole cross section fitted to
DIS data incorporates all perturbative and non-perturbative radiation
contributions.  Predictions for the LHC in the same framework are
given in Ref.~\cite{us-3}. Comparison with the predictions of other
approaches at the LHC can be found in Ref.~\cite{lhc-hic}.

\section{Colour dipole orientation}
A colorless $\bar qq$ dipole is able to interact only due to the
difference between the impact parameters of $q$ and $\bar q$ relative
to the scattering center. If $\vec s $ is the impact parameter of the
center of gravity of the dipole, and $\vec r$ is the transverse
separation of the $q$ and $\bar q$, then the azimuthal
angle of the radiated photons transverse momentum at a given impact
parameter $\vec s$ correlates with the direction of $\vec s$. 
In terms of the partial elastic amplitude
$f^{N}_{q\bar{q}}(\vec{s},\vec{r})$, it means that the vectors
$\vec{r}$ and $\vec s$ are correlated. 

One can see this in a simple example of a dipole interacting with a
quark in Born approximation. The partial elastic amplitude reads,
\beqn
\im f^q_{\bar qq}(\vec s ,\vec r) &=& \frac{2\alpha_s^{2}}{9\pi^2} \int
\frac{d^2q\,d^2q'}{(q^2+\mu^2)(q'^{2}+\mu^2)}\, 
\nonumber\\ &\times&
\left[e^{i\vec q\cdot(\vec s +\vec r/2)}- e^{i\vec q\cdot(\vec s -\vec
r/2)}\right]
\nonumber\\ &\times&
\left[e^{i\vec q^{\,\prime}\cdot(\vec s +\vec r/2)}- e^{i\vec
q^{\,\prime}\cdot(\vec s -\vec r/2)}\right]\nonumber\\
&=&  \frac{8\alpha_s^2}{9}
\left[K_0\left(\mu\left|\vec s +\frac{\vec r}{2}\right|\right) -
K_0\left(\mu\left|\vec s -\frac{\vec r}{2}\right|\right)\right]^2,
\nonumber\\
\label{e1}
 \eeqn
 where we introduced an effective gluon mass $\mu$ to take
 into account some nonperturbative effects. It is obvious from the
above expression that the partial
 elastic dipole amplitude exposes a correlation between $\vec r$ and
 $\vec s $, and the amplitude vanishes when $\vec s \cdot\vec r=0$. 
 In the above expression we assumed for the
 sake of simplicity that $q$ and $\bar q$ have equal longitudinal
 momenta, i.e. they are equally distant from the dipole center of
 gravity. The general case of unequal sharing of the dipole momentum will be
 considered later.  
 \begin{figure}[!t]
 \centerline{\includegraphics[width=5 cm] {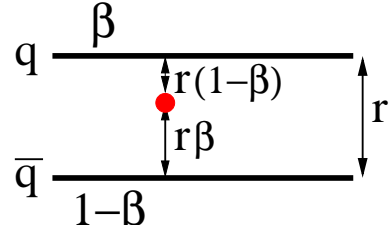}} 
\caption{The relative distance of $q$ and $\bar q$ from the center of gravity of
 $q\bar q$ dipole varies with the fractional light-cone momenta $\beta$.\label{f2}}
\end{figure}

The Born amplitude is unrealistic, since it leads to an energy independent
dipole cross section $\sigma_{\bar qq}(r,x)$. This dipole cross section
has been well probed by measurements of the proton structure function at
small Bjorken $x$ at HERA, and was found to rise towards small $x$, with
an $x$ dependent steepness.

The dipole elastic amplitude $f^{N}_{q\bar{q}}$ of a $\bar{q}q$ dipole
colliding with a proton at impact parameter $\vec s$ is given by
\cite{us-1}
 \beqn
&&\im f^N_{\bar qq}(\vec s ,\vec r,\beta)=\frac{1}{12\pi}
\int\frac{d^2q\,d^2q'}{q^2\,q'^2}\,\alpha_s\,
{\cal F}(x,\vec q,\vec q^{\,\prime})
e^{i\vec s \cdot(\vec q-\vec q^{\,\prime})}
\nonumber\\ &\times&
\left(e^{-i\vec q\cdot\vec r\beta}-
e^{i\vec q\cdot\vec r(1-\beta)}\right)\,
\left(e^{i\vec q'\cdot\vec r\beta}-
e^{-i\vec q'\cdot\vec r(1-\beta)}\right)\,
\,.
\label{300}
 \eeqn where we defined
 $\alpha_{s}=\sqrt{\alpha_{s}(q^{2})\alpha_{s}(q^{\prime 2})}$ and
 $\mathcal{F}(x,\vec{q},\vec{q}^{\,\prime})$ is the generalized
 unintegrated gluon density (see below).  The $x$ dependence is
 implicit in the above expression.  The fractional light-cone momenta
 of the quark and antiquark are denoted by $\beta$ and $1-\beta$,
 respectively.  It is obvious that the center of gravity of $q\bar{q}$
 is closer to the fastest $q$ or $\bar{q}$, see Fig.~(\ref{f2}). The
 radiated photon takes away a fraction $\alpha$ of the quark momentum,
 see Fig.~(\ref{f1}). Therefore, for photon production, we have
\beq
\beta=\frac{1}{2-\alpha}. \label{kin}
 \eeq

Integrating over the vector $\vec s$ one can recover the dipole cross
section $\sigma^{N}_{q\bar{q}}(r,x)$. 
\begin{eqnarray}
\sigma^{N}_{q\bar{q}}(r,x)&=&2\int d^{2}\vec s~\text{Im}f^{N}_{q\bar{q}}(\vec s,\vec{r},\beta)\nonumber\\
&=&\frac{4\pi}{3}\int\frac{d^{2}q}{q^{4}}(1-e^{-i\vec{q}.\vec{r}})\alpha_{s}(q^{2})\mathcal{F}(x,q).
\label{di-app}\
\end{eqnarray}
It is important to notice that the expression Eq.~(\ref{300}) also goes 
beyond the usual assumption that the dipole cross section is independent 
of the light-cone momentum sharing $\beta$. Although, the partial amplitude 
Eq.~(\ref{300}) does depend on $\beta$, this dependence disappears 
after integration over impact parameter $\vec s $ as shown in Eq.~(\ref{di-app}).

 The generalized unintegrated gluon density
$\mathcal{F}(x,\vec{q},\vec{q}^{\,\prime})$ is related to the diagonal one
by
 \beq
\mathcal{F}(x,\vec{q},\vec{q}^{\,\prime}=\vec{q}) =
\mathcal{F}(x,q).
\eeq
The generalized unintegrated gluon density in Born approximation takes the
form,
 \beqn
{\cal F}(x,\vec q,\vec q^{\,\prime}) &\Rightarrow&
{\cal F}_{Born}(\vec q,\vec q^{\,\prime})
\nonumber\\ &=&
\frac{4\alpha_s}{\pi}
\left[F_N(\vec q-\vec q^{\,\prime})-
F_N^{(2q)}(\vec q,\vec q^{\,\prime})\right],
\label{310}
 \eeqn
where $F_N(k)=\la\Psi_N|\exp(i\vec k\cdot\vec\rho_1)|\Psi_N\ra$ is the
nucleon form factor, and $F_N^{(2q)}(\vec q,\vec
q^{\,\prime})=\la\Psi_N|\exp[i\vec q\cdot\vec\rho_1-i\vec
q^{\,\prime}\cdot\vec\rho_2]|\Psi_N\ra$ is the so called two-quark nucleon
form factor which can be calculated using the three valence quark nucleon wave
function $\Psi_N(\vec\rho_1,\vec\rho_2,\vec\rho_3)$.
 \begin{figure}[!t]
       \centerline{\includegraphics[width=8 cm] {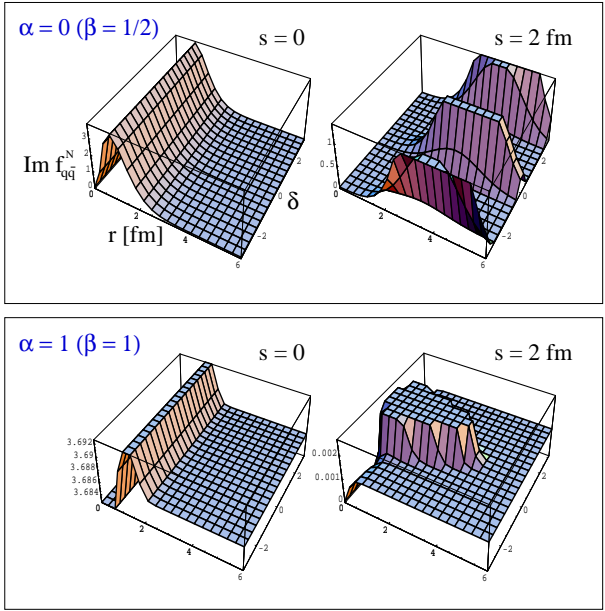}}
       \caption{  The partial elastic amplitude
       $\text{Im}f^{N}_{q\bar{q}}$ (mb) of the $\bar{q}q$ dipole on a
       proton at impact parameter $s$ as a function of dipole size $r$
       and angle $\delta$ between $\vec{s}$ and $\vec{r}$ for two values of $\alpha=0, 1$. We
       use a fixed value of $x=0.01$ for all
       plots. \label{f3}}
\end{figure}

For the dipole cross section we rely on the popular saturated shape
\cite{kst} fitted to HERA data for $F_2^p(x,Q^2)$. Assuming no correlation 
between the momenta $\vec q$ and $\vec q^{\,\prime}$ inside the
Pomeron aside from the Pomeron-proton form factor, we arrive at the
following form of ${\cal F}(x,\vec q,\vec q^{\,\prime})$ \cite{us-1},
\beqn {\cal
F}(x,\vec q,\vec q^{\,\prime}) &=&
\frac{3\,\sigma_0}{16\,\pi^2\,\alpha_s}\ q^2\,q'^2\,R_0^2(x)
\nonumber\\ &\times&
{\rm exp}\Bigl[-{1\over8}\,R_0^2(x)\,(q^2+q'^2)\Bigr]
\nonumber\\ &\times&
{\rm exp}\bigl[-R_N^2(\vec q-\vec q^{\,\prime})^2/2\bigr]
\,,
 \label{320} \eeqn where $\sigma_{0}=23.03$ mb, $R_{0}(x)=0.4 \fm
 \times (x/x_{0})^{0.144}$ with $x_{0}=3.04\times 10^{-4}$ \cite{kst}.
 We assume here that the Pomeron-proton form factor has the Gaussian
 form, $F^p_\Pom(k_T^2)=\exp(-k_T^2 R_N^2/2)$, so the slope of the
 $pp$ elastic differential cross section is
 $B^{pp}_{el}=2R_N^2+2\alpha^\prime_\Pom\ln(s/s_0)$, where
 $\alpha^\prime_\Pom\approx 0.25\GeV^{-2}$ is the slope of the Pomeron
 trajectory, $s_0=1\GeV^2$. $R_N^2\approx \la r_{ch}^2\ra/3$ is the
 part of the slope of elastic cross section related to the
 Pomeron-proton form factor and $\la r_{ch}^2\ra$ is the mean-square
 charge radius of the proton.

Unfortunately, it is not possible to uniquely determine the
unintegrated gluon density function from the available
data. Nevertheless, the proposed form Eq.~(\ref{320}) seems to be a
natural generalization which preserves the saturation properties of
the diagonal part \cite{us-1}.

With this unintegrated gluon density the partial amplitude Eq.~(\ref{300})
can be calculated explicitly,
\begin{widetext}
 \beq
\im f^N_{\bar qq}(\vec s ,\vec r,x,\beta) =
\frac{\sigma_0}{8\pi B_{el}}\,
\Biggl\{\exp\left[-\frac{[\vec s +\vec r(1-\beta)]^2}{2B_{el}}\right] +
\exp\left[-\frac{(\vec s -\vec r\beta)^2}{2B_{el}}\right]
-2\exp\Biggl[-\frac{r^2}{R_0^2}-
\frac{[\vec s +(1/2-\beta)\vec r]^2}{2B_{el}}\Biggr]
\Biggr\},
\label{340}
 \eeq
\end{widetext}
 where $B_{el}(x)=R_N^2+R_0^2(x)/8$.

In Fig.~(\ref{f3}) we show the partial dipole amplitude
$f^{N}_{q\bar{q}}(\vec s,\vec{r})$ as a function of the dipole size
$r$ and the angle $\delta$ between $\vec s$ and $\vec{r}$, at various
fixed values of $s$, for two values of $\alpha=0, 1$.  One can see that
for very small dipole sizes $r$ the dipole orientation is not
important. For very large dipole sizes $r$ compared to the impact
parameter $s$ or very small values of $s$ the dipole orientation is
also not present. It is important to note that the generic feature of
the partial dipole amplitude, e. g. its maximum and minimum pattern,
changes with $\alpha$.  However, it is not obvious a priori how the convolution
with the proton structure function Eq.~(\ref{con1}), which leads to a sum over all different configurations of 
$\alpha$ and on top of that the convolution between the partial dipole
amplitude and the nuclear profile Eq.~(\ref{eik}), which leads to a even more
complicated angle mixing, gives rise to a final azimuthal asymmetry.
\begin{figure}[!t]
       \centerline{\includegraphics[width=8 cm] {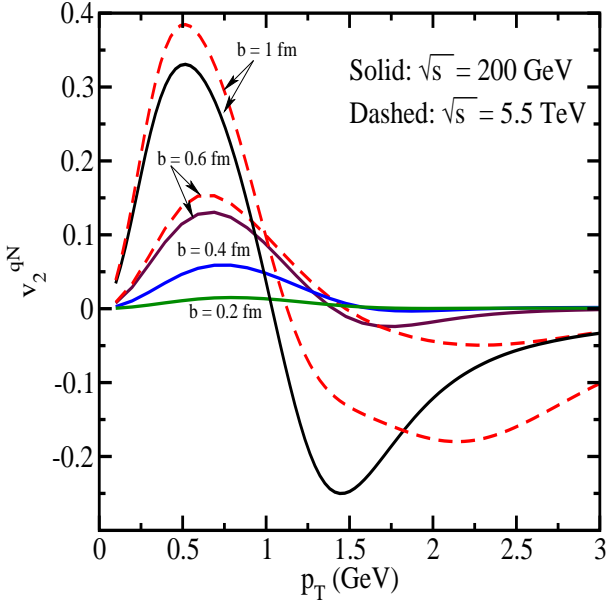}}
       \caption{The anisotropy parameter $v_2^{qN}(b,p,\alpha)$ as function of $p_T$
calculated at $\alpha=1$ for different impact parameters $b$ and energies:
$\sqrt{s}=200\GeV$ (solid, $b=0.2,\ 0.4,\ 0.6,\ 1\fm$), and
$\sqrt{s}=5500\GeV$ (dashed, $b=0.6,\ 1\fm$). The plot is taken from Ref.~\cite{us-1}. \label{vqn}}
\end{figure}

\section{Azimuthal asymmetry}
The main source of azimuthal asymmetry in the amplitude (\ref{eik})
is the interplay between multiple rescattering and the shape of the
physical system.  The key function which describes the effect of
multiple interactions is the eikonal exponential in Eq.~(\ref{eik}),
while the information about the shape of the system is incorporated
through a convolution of the impact parameter dependent partial
elastic amplitude and the nuclear thickness function. Notice that
the initial space-time asymmetry gets translated into a momentum
space anisotropy by the double Fourier transform in Eq.~(\ref{m1}).

The azimuthal asymmetry of prompt photon production, resulting
from parton-nucleus (qt) or proton-nucleus (pt) collisions for $t=N, A$, is defined as the second order
Fourier coefficients in a Fourier expansion of the azimuthal
dependence of a single-particle spectra Eq.~({\ref{m1}) around the
beam direction,
\begin{equation}
v_{2}^{q(p)t}(p_{T},b,
\alpha)= \frac{\int_{-\pi}^{\pi} d\phi \cos(2\phi)
\frac{ d\sigma^{\gamma}(q(p)t\to\gamma X)}{d(ln \alpha)d^{2}\vec{p}_{T}d^{2}\vec{b}}}
{\int_{-\pi}^{\pi} d\phi \frac{ d\sigma^{\gamma}(q(p)t\to\gamma X)}{d(ln
\alpha)d^{2}\vec{p}_{T}d^{2}\vec{b}}}, \label{v2-1}
\end{equation}
where the angle $\phi$ is defined with respect to the reaction
plane. In the same fashion, the
azimuthal asymmetry of photon yield from collisions of two nucleus
A$_{1}$ and A$_{2}$ at impact parameter $B$ is defined as
\begin{eqnarray}
v_{2}^{A_{1}A_{2}}(B,p_{T})&=&
\frac{\int_{-\pi}^{\pi} d\phi \cos(2\phi)~\mathcal{G}_{N}}
{\int_{-\pi}^{\pi} d\phi~\mathcal{G}_{D}};\nonumber\\
\mathcal{G}_{N}&=& \int d^{2}\vec{b}
\cos(2\Theta_{1}) \frac{d\sigma^{\gamma}(pA_{1}\to \gamma
X)}{dx_{F}d^{2}\vec{p}_{T}d^{2}\vec{b}_{1}} T_{A_{2}}(\vec{b}_{2})\nonumber\\
&+& \int d^{2}\vec{b}\cos(2\Theta_{2}) \frac{d\sigma^{\gamma}(pA_{2}\to \gamma
X)}{dx_{F}d^{2}\vec{p}_{T}d^{2}\vec{b}_{2}} T_{A_{1}}(\vec{b}_{1});
\nonumber\\
\mathcal{G}_{D}&=&\int d^{2}\vec{b} \frac{d \sigma^{\gamma}(pA_{1}\to \gamma
X)}{dx_{F}d^{2}\vec{p}_{T}d^{2}\vec{b}_{1}} T_{A_{2}}(\vec{b}_{2})\nonumber\\
&+&\int d^{2}\vec{b}\frac{d \sigma^{\gamma}(pA_{2}\to \gamma X)}{dx_{F}d^{2}\vec{p}_{T}d^{2}\vec{b}_{2}} T_{A_{1}}(\vec{b}_{1}),
\label{v2-3}\
\end{eqnarray}
 where we used the notation $\vec{b}_{2}=\vec{b}+\vec{B}$,
 $\vec{b}_{1}=\vec{b}$ ($\vec{b}$ is the impact parameter of the
 p$A_{1}$ collision) and the angle $\Theta_{1}$ ($\Theta_{2}$) is the
 angle between the vectors $\vec{b}_{1}$( $\vec{b}_{2}$) and
 $\vec{B}$, respectively. The medium modification of nucleon structure
 functions in our interested range of $p_T$ is less than $20\%$ and is
 ignored in the above expression.

The only external input in our approach is the nuclear profile. First,
we take a popular Woods-Saxon (WS) profile, with a nuclear radius
$R_{A}=6.5$ fm and a surface thickness $\xi=0.54$ fm, for Pb+Pb
collisions \cite{ws}.

In Fig.~\ref{vqn}, we show examples of azimuthal anisotropy from 
quark-nucleon collisions radiating a photon $v_2^{qN}(b,p,\alpha)$, with $\alpha=1$ and at different impact
parameters and energies. The results show that the anisotropy of the dipole interaction rises with
impact parameter, reaching rather large values. As function of the transverse
momentum of the radiated photons, $v_2^{qN}(b,p,\alpha)$ vanishes at large
$p_T$. Such a behavior could be anticipated, since the interaction of
vanishingly small dipoles responsible for large $p$ is not sensitive to the
dipole orientation.
\begin{figure}[!t]
       \centerline{\includegraphics[width=8 cm] {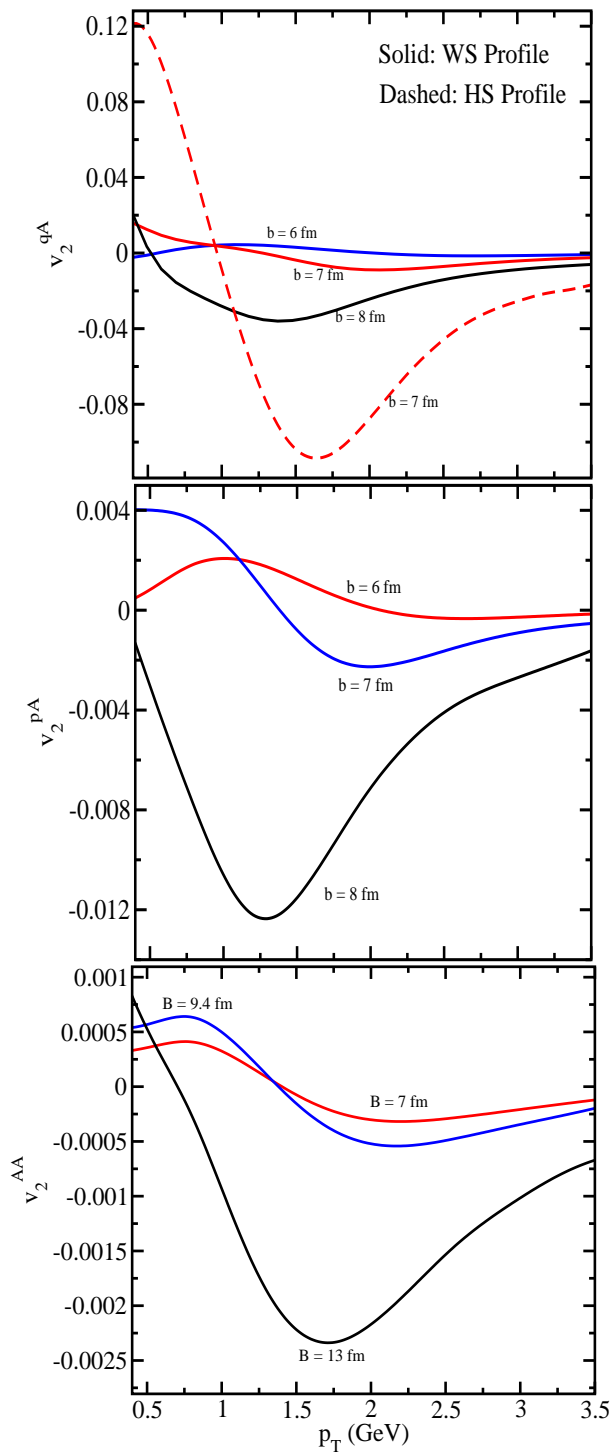}}
       \caption{The impact parameter dependence of prompt photon
       azimuthal asymmetry, for q+Pb (at $\alpha=1$), p+Pb and Pb+Pb collisions for
       RHIC energy at midrapidities, with the Woods-Saxon (WS) nuclear profile. As example, we have
       also shown in the first panel, the prompt photon elliptic
       anisotropy from q+pb collisions for the hard sphere (HS) nuclear
       profile. The plot is taken from Ref.~\cite{us-2}. \label{vaa}}
\end{figure}

In Fig.~(\ref{vaa}), we show the calculated values of $v_{2}^{q(p)A}$
defined in Eq.~(\ref{v2-1}), for fixed $\alpha=1$,  at various q(p)A
collision impact parameters $b$ for the RHIC energy $\sqrt{\bold{s}}=200$ GeV at midrapidities.  If the nuclear profile
function was constant, then the convolution between the nuclear
profile and the dipole orientation, defined in Eq.~(\ref{eik}),
would be trivial, and $v_{2}^{qA}$ becomes then identically zero.
Therefore, the main source of azimuthal anisotropy is not present
for central collisions where the correlation between nuclear profile and
dipole orientation is minimal. This can be seen in Fig.~(\ref{vaa}),
where a pronounced elliptic anisotropy is observed for collisions
with impact parameters close to the nuclear radius $R_{A}$, where
the nuclear profile undergoes rapid changes. Therefore, the important
parameter which controls the elliptic asymmetry in this mechanism is
$|b-R_{A}|$ \cite{us-1}.

It is important to notice that $v^{qA}_2$ is suppressed an order of
magnitude compared to $v^{qN}_2$. At first glance this might look
strange, since the quark interacts with nucleons anyway. However, a
quark propagating through a nucleus interacts with different nucleons
located at different azimuthal angles relative to the quark
trajectory. Their contributions to $v_2^{qA}$ tend to cancel each
other, restoring the azimuthal symmetry. Such cancellation would be
exact if the nuclear profile function $T_A(b)$ were constant. We have
a nonzero, but small $v_2^{qA}$ only due to the variation of $T_A$
with $b$. Going from $q(p)A$ to $AA$ collisions, see Fig.~(\ref{vaa}), the elliptic
asymmetry is further reduced. 
The main reason is that the
integrand in Eq.~(\ref{v2-3}) gets contributions only from
semi-peripheral pA collisions where our mechanism is at work, and
most of the integral over $\vec{b}$ does not contribute. This
significantly dilutes the signal. 

\section{On the sign of $\bold{v_2}$}
In Fig.~(\ref{f3}), we showed that the general behaviour of dipole
amplitude orientation, e.g. its maximum and minimum pattern changes
with the parameter $\beta$ which defines the relative position of the
center of gravity of the dipole from the $q$ and $\bar q$, see
Fig.~(\ref{f2}).  Here, we take a heuristic approach, and explore a
possible link between the peculiar behaviour of $v_2$ shown in Fig.~(\ref{vaa})
with the dipole orientation introduced in Eq.~(\ref{300}). 

Let us assume for sake of argument that the parameter $\beta$ is
independent of $\alpha$ and assume that $q$ and $\bar q$ have equal
longitudinal momenta namely $\beta=1/2$.  This corresponds to a
particular configuration in which the dipole amplitude is symmetric
under $\vec r\to -\vec r$, see Fig.~(\ref{f3}). In principle, this
configuration is kinematically less probable for direct photon
production (in contrast to DIS) since it corresponds to $\alpha=0$ via
Eq.~(\ref{kin}). Notice that although we take a fixed $\beta=1/2$ in the
dipole amplitude Eq.~(\ref{300}), the LC distribution of the
projectile quark $\gamma q$ fluctuation Eq.~(\ref{dylcl}) still
depends on $\alpha$ and also the transverse dipole size is $\alpha \vec
r$ and varies with $\alpha$, see Eq.~(\ref{m1}).

We repeat the computation of the azimuthal asymmetry $v_2$ of prompt
photons in the same way as discussed in the previous section.  For
example, in Fig.~(\ref{vaa1}), we show the anisotropy asymmetry $v_{2}^{qN}$ and
$v_{2}^{qA}$ at $\alpha=1$ for various impact parameters as in Figs.~(\ref{vqn},\ref{vaa}). Comparing
with the results presented in the previous section, it is seen that
although the order of magnitude of $v_2$ is the same in both cases, now the
sign of $v_2$ does not change at higher $p_T$ and remains positive.
This indicates that the sign of $v_2$ in this mechanism is related to the dipole orientation via 
the parameter $\beta$.   

Notice also that the sign behaviour of the prompt photon
$v_2$ for $AA$ collisions at higher $p_T$ is also present for both
$qN$ and $qA$ collisions.

\begin{figure}[!t]
       \centerline{\includegraphics[width=8 cm] {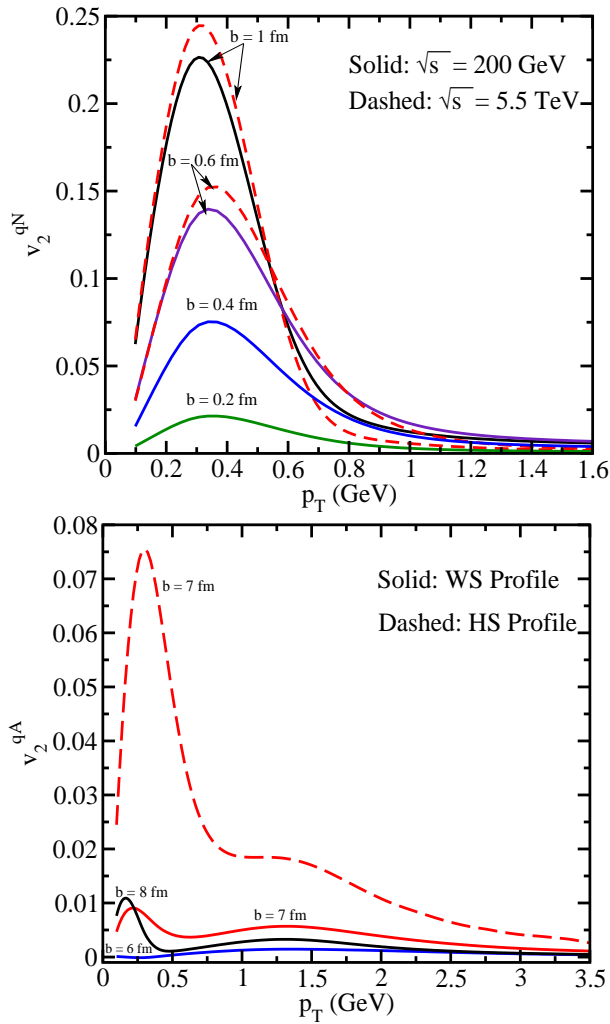}}
       \caption{ Upper panel: The anisotropy parameter
       $v_2^{qN}(b,p,\alpha)$ as function of $p_T$ for different
       impact parameters $b$ for RHIC and LHC energies. Down panel:
       The anisotropy parameter $v_2^{qA}(b,p,\alpha)$ as function of
       $p_T$ for various impact parameter $b$ for RHIC energy at midrapidity. Similar to
       Fig.~(\ref{vaa}) we show the results for both the WS and the HS
       nuclear profiles. In both plots the results calculated at a fixed
       $\beta=1/2$ in the dipole amplitude and
       $\alpha=1$. \label{vaa1}}
\end{figure}

\section{Concluding remarks}
 The azimuthal elliptic asymmetry $v_2$ observed in heavy ion collisions 
is usually associated with properties of the medium created in the final
state. We introduced a novel mechanism which relates this azimuthal asymmetry
to the colour dipole orientation. To this end, we proposed a model
generalizing the unintegrated gluon density fitted to data for the proton
structure function to an off-diagonal unintegrated gluon distribution.  

We showed that the azimuthal asymmetry $v_2$ of prompt photons changes sign and becomes negative for peripheral collisions.  Although this
behaviour seems to be robust for the considered range of $p_{T}$, there is some
uncertainty on the magnitude of the $v2$ coming from this mechanism. 
The shape of the tail of nuclear
profile is important in this mechanism, since it significantly affects the
results. To highlight this point, in Figs.~(\ref{vaa},\ref{vaa1}), we have
also shown the azimuthal asymmetry from quark-nucleus collisions
$v_{2}^{qA}$ for the hard sphere (HS) nuclear profile. By comparing with
the results from the WS nuclear profile for the same setting, one may
conclude that the maximum uncertainty in this mechanism can be as big as
an order of magnitude. Unfortunately the tail of all available nuclear
profile parametrizations is less reliable and obtained by a simple
extrapolation \cite{ws}. This is also due to the fact that the neutron
distribution, which may be more important on the periphery, cannot be
properly accounted for by electron scattering data.  Another source of
uncertainty in this approach is due to the fact that the off-diagonal part
of the unintegrated gluon density cannot be uniquely defined from the
current experimental data.

In order to see if this mechanism is relevant to heavy ion
collisions, it would be of great interest to calculate the azimuthal
asymmetry for gluon radiation and hadron production at RHIC. This
mechanism might also contribute to the azimuthal asymmetry in DIS and in
the production of dileptons. We plan to report on some of these
problems in the near future.

 \begin{acknowledgments}

One of us (AHR) would like to thank the organizers of ``{\em II Latin
American Workshop on High Energy Phenomenology}'' for the kind invitation
and for the very interesting and stimulating meeting. This work was
supported in part by Conicyt (Chile) Programa Bicentenario PSD-91-2006, by
Fondecyt (Chile) grants 1070517 and 1050589, and by DFG (Germany) grant 
PI182/3-1.

\end{acknowledgments}

\end{document}